\documentclass{article}%
\usepackage{amsfonts}
\usepackage{amssymb}
\usepackage{amsmath}
\usepackage{amsthm}
\usepackage{fullpage}
\usepackage{xcolor}
\usepackage{graphicx}
\usepackage{graphics}
\usepackage{pstricks}
\usepackage{pst-node}
\usepackage{pst-plot}
\usepackage{slashbox}
\usepackage{float}%
\providecommand{\U}[1]{\protect\rule{.1in}{.1in}}
\allowdisplaybreaks

\newtheorem{example}{Example}

\begin{document}

\title{An Encoding-Decoding algorithm based on 
Padovan numbers}

\author{Jenan Shtayat, Alaa Al-Kateeb\footnote{
Yarmouk university, Irbid-Jordan.}}

\maketitle

\begin{abstract}
In this paper, we propose a new  of coding/decoding algorithm using Padovan
Q-matrices. This method is based on  blocked message matrices. an advantage
of this method is that the  encryption of each message matrix will use a different key and that will increase the security of the method. 
\end{abstract}

\textbf{keywords:} ~~Padovan sequence; Padovan Q-matrix; Encoding-Decoding algorithms.


\section{Introduction}  The Padovan sequence is the sequence of integers $P_n$ defined
by the initial values $P_0 = 0, P_1 = 0, P_2 = 1$ and the recurrence relation \[P_n=P_{n-2}+P_{n-3} ~~~~ n  \geq 3\]

 The Padovan sequence  is named after Richard Padovan who attributed its discovery to Dutch architect Hans van der Laan in his 1994 essay Dom. Hans van der Laan : Modern Primitive, see \cite{wiki}, the first few terms of the sequence are: $0, 0, 1, 0, 1, 1, 1, 2, 2, 3, 4, 5, 7, 9, 12, 16,
21, 28, 37, 49, 65, \dots$.\\

The Padovan numbers have the   $Q-$matrix, where  $Q=\begin{bmatrix}0 & 1 & 0 \\
0 & 0 & 1 \\
1 & 1 & 0 \\
\end{bmatrix}$, clearly\\ $\det(Q)=1$.  It is well-known that \cite{SO1}
 \[Q^n=\begin{bmatrix}P_{n-1}& P_{n+1} & P_n \\
P_n & P_{n+2} & P_{n+1} \\
P_{n+1} & P_{n+3} & P_{n+2} \\
\end{bmatrix}\]
Recently, coding-decoding algorithms becomes very important for improving information security and data transfers. Especially Fibonacci coding \cite{ST1,ST2,TU}.\\

  In this paper we  propose a new coding-decoding algorithm using the  Padovan $Q-$matrix, the algorithm depends on dividing the message matrix
into the block matrices each is of size $3 \times 3$. We use different numbered alphabet for each
message, so we get a more reliable coding method. The alphabet is determined by
the number of block matrices of the message matrix.  \\

This paper is structured as follows: in section \ref{sec:2} we describe the new code depends on the Padovan numbers, in section \ref{sec:3} we list two detailed examples, and finally in section \ref{sec:4} we give a Maple code that can be used to check the computations.  
\section{The Encoding and Decoding  Algorithms}\label{sec:2} 
At first, we put our message in a matrix  $M$ of  size $3m \times 3m$ adding commas between two words. Dividing the message square matrix $M$ of size $3m  \times3m$ into the matrices, named
$B_i$  of size $3 \times 3$.
For simplicity and readability  for $1\leq i\leq m^2 $, let 
\begin{enumerate}

\item  $B_i=\begin{bmatrix}b_1^i & b_2^i & b_3^i \\
b_4^i & b_5^i & b_6^i \\
b_7^i & b_8^i & b_9^i \\
\end{bmatrix}$ 
\item  $ E_i=\begin{bmatrix}e_1^i & e_2^i & e_3^i \\
e_4^i & e_5^i & e_6^i \\
e_7^i & e_8^i & e_9^i \\
\end{bmatrix}$
\item  $Q^n=\begin{bmatrix}q_1 & q_2 & q_3 \\
q_4 & q_5 & q_6 \\
q_7 & q_8 & q_9 \\
\end{bmatrix}$ 

\end{enumerate}
\subsection{Coding Algorithm}
Before applying the algorithm we need to make sure that in the matrices $B_i$, the minor of the entries $(B_i)_{22}$ is not zero, otherwise we change the message by adding a sequence of  zeros    in the begging of the message (and  in the end we can do that as many times as needed). 
\begin{enumerate}
\item  We choose $n$ as follows:
\[n=\begin{cases}4 & m =1 \\
m^2 & otherwise\\
\end{cases}\]
\item We use the following table $\mod 28$ \[
 \begin{tabular}{|c|c|c|c|c|c|c|c|c|c|}\hline
A & B & C & D & E & F & G & H & I & J \\\hline
$n$ & $n+1$ &  $n+2$& $n+3$ & $n+4$ &$n+5$  & $n+6$ &  $n+7$&  $n+8$& $n+9$\\\hline
K & L & M&N & O & P & Q &R  & S &T  \\\hline
$n+10$ & $n+11$ & $n+12$ & $n+13$ & $n+14$ & $n+15$ & $n+16$ &$n+17$  &$n+18$  &$n+19$  \\\hline
 U&V&  W&  X& Y &Z  & , &0 &  &   \\\hline
 $n+20$&$n+21$  & $n+22$ & $n+23$ & $n+24$ &  $n+25$& $n+26$ & $n+27$ &  &  \\\hline
\end{tabular}\]
\item Compute $d_i=\det(B_i)$.
\item Construct the  coded matrix  $C=[d_i,b_k^i]_{k\in \{1,2,3,4,6,7,8,9\}}$
 \end{enumerate}
\subsection{ Decoding Algorithm}
we have $\det(Q^n B_i)=d_i$
\begin{enumerate}
\item
Compute
 \[\begin{array}{|c|c|c|c|c|c|}
\hline
e_1^i & q_1b_1^i+q_2b_4^i+q_3b_7^i & e_4^i & q_4b_1^i+q_5b_4^i+q_6b_7^i & e_7^i & q_7b_1^i+q_8b_4^i+q_9b_7^i \\\hline
e_2^i & q_1b_2^i+q_3b_8^i & e_5^i & q_4b_2^i+q_6b_8^i  & e_8^i & q_7b_2^i+q_9b_8^i \\\hline
e_3^i& q_1b_3^i+q_2b_6^i+q_3b_9^i & e_6^i & q_4b_3^i+q_5b_6^i+q_6b_9^i
 & e_9^i & q_7b_3^i+q_8b_6^i+q_9b_9^i \\\hline
\end{array}\]

\item Solve the equation \[d_i=(e_2+q_2x)\begin{vmatrix}e_6 & e_4 \\
e_9 & e_7 \\
\end{vmatrix}+(e_5+q_5x)\begin{vmatrix}e_7 & e_1 \\
e_9 & e_3 \\
\end{vmatrix}+(e_8+q_8x)\begin{vmatrix}e_3 & e_1 \\
e_6 & e_4 \\
\end{vmatrix}
              \]
\item Substitute for $x_i=b_5^i$
\item Construct $B_i$
\item Construct M
\end{enumerate}

\section{Examples}\label{sec:3}
\begin{example}[A short example] Consider the message \textquotedblleft HELLOALA"  \[M=\begin{bmatrix}H & E & L \\
L & O & , \\
A & L & A \\
\end{bmatrix}\]
\begin{description}
\item[Encoding:] Since $m=1$, we have one block $B_1=M$.  
\begin{description}
\item[Step 1]   $n=4$, hence $B_1=\begin{bmatrix}11& 8 & 15 \\
15 & 18 & 3 \\
4 & 15 & 4 \\
\end{bmatrix}$ 
\item[Step 2] $d_1=2208$
\item  [Step 3]we have 
 $C=[2208,11,8,15,15,3,4,15,4]$
\end{description}
\item [Decoding:] We have
\begin{description} 
\item[Step 1]   $Q^4=\begin{bmatrix}0 & 1 & 1 \\
1 & 1 & 1 \\
1 & 2 & 1 \\
\end{bmatrix}$
\item [Step 2] \[\begin{array}{|c|c|c|c|c|c|}
\hline
e_1^1 & 19 & e_4^1 & 30 & e_7^1 & 45 \\\hline
e_2^1 & 15 & e_5^1 & 23  & e_8^i & 23\\\hline
e_3^1& 7 & e_6^1 & 22 & e_9^i & 25 \\\hline
\end{array}\]
\item [Step 3 ]Solve $2208=2496-16x_1$

%\end{align*}
\item[Step 4] $x_1=18$ and $b_5^i=18$  
\item [Step 5]$B_1=\begin{bmatrix}11& 8 & 15 \\
15 & 18 & 3 \\
4 & 15 & 4 \\
\end{bmatrix}$ 
\end{description}
\end{description} \end{example}
\begin{example}[A long example] Consider the message \textquotedblleft HELLOTOBETHEBESTDOYOURBEST\textquotedblright
\[M=\begin{bmatrix}

H & E & L & L & O & , \\
T & O & , & B & E & , \\
T & H & E & , & B & E \\
S & T & ,& D & O& , \\
Y & O & U & R & , & B \\
E & S & T & , & , & , \\
\end{bmatrix}\]
\begin{description}
\item[Encoding:] We have the following steps
\begin{description}
\item[Step 1]
  Since $m=2$, we We  divide the message matrix into  four block matrices.\\ $B_1=\begin{bmatrix}H & E & L \\
L & O & , \\
T & O & , \\
\end{bmatrix}$, $B_2=\begin{bmatrix}B & E & , \\
T & H & E \\
, & B & E \\
\end{bmatrix}$, $B_3=\begin{bmatrix}S & T & , \\
D & O & , \\
Y & O & U \\
\end{bmatrix}$, $B_4=\begin{bmatrix}R & , & B \\
E & S & T \\
, & , & , \\
\end{bmatrix}$    
%\begin{enumerate}
\item[Step 2]\begin{enumerate}
\item 
 $B_1=\begin{bmatrix}11& 8 &15  \\
15 & 18 & 3 \\
23 &18  &3  \\
\end{bmatrix}$ we have $d_1=-1968$
\item  $B_2=\begin{bmatrix}5&  8& 3 \\
 23& 11 &8  \\
3 &5  &8  \\
\end{bmatrix}$ we have  $d_2=-794$
\item  $B_3=\begin{bmatrix}22& 23 &3  \\
7 & 18 & 3 \\
1 &18  &24  \\
\end{bmatrix}$ we have $d_3=4845$
\item  $B_4=\begin{bmatrix}21&  3& 5 \\
 8& 22 &23 \\
3 &3  &3  \\
\end{bmatrix}$ we have $d_4=-138$\end{enumerate} 
%\item $d_1=2208$
%\item  we have 
 %$C=[2208,11,8,15,15,3,4,15,4]$
\item[Step 3] we get $C=\begin{bmatrix}-1968 & 11 & 8 & 15 & 15 & 3 & 23 & 18 & 3 \\
-794 & 5 & 8 & 3 & 23 & 8 & 3 & 5 & 8 \\
4845 & 22 & 23 & 3 & 7 & 3 & 1 & 18 & 24 \\
-138 & 21 & 3 & 5 & 8 & 23 & 3 & 3 & 3 \\
\end{bmatrix}$
\end{description}
\item [Decoding:]\
\begin{enumerate}
\item For $B_1$\begin{description}
\item 
[Step1]\[\begin{array}{|c|c|c|c|c|c|}
\hline
e_1^1 & 38 & e_4^1 & 49& e_7^1 & 64 \\\hline
e_2^1 & 18 & e_5^1 & 26  & e_8^i & 26\\\hline
e_3^1& 6 & e_6^1 & 21 & e_9^i & 24\\\hline
\end{array}\]
\item[Step 2] Solve $-138=48x_1-1194$
\item [Step 3] $x_1=22$
\end{description}
\item For $B_2$\begin{description}
\item 
[Step1]\[\begin{array}{|c|c|c|c|c|c|}
\hline
e_1^2 & 26 & e_4^2 & 31& e_7^2 & 54 \\\hline
e_2^2 & 5 & e_5^2 & 13  & e_8^2 & 13\\\hline
e_3^2& 16 & e_6^2 & 19 & e_9^2 & 27\\\hline
\end{array}\]
\item[Step 2] Solve $-794=31x_2-1135$
\item [Step 3] $x_2=11$
\end{description}
\item For $B_3$\begin{description}
\item 
[Step1]\[\begin{array}{|c|c|c|c|c|c|}
\hline
e_1^3 & 8 & e_4^3 & 30& e_7^3 & 37 \\\hline
e_2^3 & 18 & e_5^3 & 41& e_8^3 & 41\\\hline
e_3^3& 27 & e_6^3 & 30 & e_9^3 & 33\\\hline
\end{array}\]
\item[Step 2] Solve $4845=525x_3-4605$
\item[Step 3] $x_3=18$
\end{description}
\item For $B_4$\begin{description}
\item 
[Step1]\[\begin{array}{|c|c|c|c|c|c|}
\hline
e_1^4 & 11 & e_4^4 & 32& e_7^4 & 40 \\\hline
e_2^4 & 3 & e_5^4 & 6 & e_8^4 & 6\\\hline
e_3^4& 26 & e_6^4 & 31 & e_9^4 & 54\\\hline
\end{array}\]
\item[Step 2] Solve $-138=48x_4-1194$
\item[Step 3] $x_4=22$
\end{description}

\end{enumerate}
\end{description}
\end{example}
\begin{example}[With minor equals zero]Consider the message \textquotedblleft ALAJENAN"
 \[M=\begin{bmatrix} A& L & A \\
, & J & E \\
N & A & N \\
\end{bmatrix}\]Here we vane only one block matrix \\
$B_1=\begin{bmatrix}4&15  & 4 \\
26 & 13 & 8 \\
17 & 4 & 17 \\
\end{bmatrix}$. Notice that the minor $M_{22}=0$, so we change the message to be \[M=\begin{bmatrix}27& 4 & 15 & 4 & 26 & 13 \\
8 & 17 & 4 & 17 & 27 & 27\\
27 & 27 & 27 & 27 & 27 & 27 \\
27& 27 & 27& 27 & 27 & 27 \\
27 & 27 & 27 & 27 & 27 & 27 \\
27 & 27 & 27 & 27 & 27 & 27 \\
\end{bmatrix}\]
\begin{description}
\item[Encoding:] We have the following steps?? fix everything??
\begin{description}
\item[Step 1]
  Since $m=2$, we We  divide the message matrix into  four block matrices.\\ $B_1=\begin{bmatrix}27 & 4 & 15 \\
8 & 17 & 4 \\
27 & 27 & 27 \\
\end{bmatrix}$, $B_2=\begin{bmatrix}4 & 26& 13 \\
17 & 27 & 27 \\
27 & 27 & 27 \\
\end{bmatrix}$, $B_3=\begin{bmatrix}27 & 27 & 27 \\
27 & 27 & 27 \\
27 & 27 & 27 \\
\end{bmatrix}$, $B_4=\begin{bmatrix}27 & 27 & 27 \\
27 & 27 & 27 \\
27 & 27 & 27 \\
\end{bmatrix}$    
%\begin{enumerate}
\item[Step 2]\begin{enumerate}
\item 
$B_1=\begin{bmatrix}27 & 4 & 15 \\
8 & 17 & 4 \\
27 & 27 & 27 \\
\end{bmatrix}$ we have $d_1=5400$
\item  $B_2=\begin{bmatrix}4&  26& 13 \\
 17& 27 &27  \\
27 &27  &27  \\
\end{bmatrix}$ we have  $d_2=3510$
\item  $B_3=\begin{bmatrix}27& 27 &27  \\
27 & 27 & 27 \\
27 &27  &27  \\
\end{bmatrix}$ we have $d_3=0$
\item  $B_4=\begin{bmatrix}27&  27& 27 \\
 27& 27 &27 \\
27 &27  &27  \\
\end{bmatrix}$ we have $d_4=0$\end{enumerate} 
%\item $d_1=2208$
%\item  we have 
 %$C=[2208,11,8,15,15,3,4,15,4]$
\item[Step 3] we get $C=\begin{bmatrix}4500 & 27 &4  &15  &8  &4  &27  &27  &27  \\
 3510& 4 & 26 & 13 & 17 &27  &27  27&27  &27  \\
0 & 27 & 27 & 27 & 27 & 27 & 27 & 27 &27  \\
0 & 27 & 27 & 27 & 27 & 27 & 27 & 27 &27  \\
\end{bmatrix}$
\end{description}
\item [Decoding:]\
\begin{enumerate}
\item For $B_1$\begin{description}
\item 
[Step1]\[\begin{array}{|c|c|c|c|c|c|}
\hline
e_1^1 & 35 & e_4^1 & 62& e_7^1 & 70 \\\hline
e_2^1 & 27 & e_5^1 & 31  & e_8^1 & 31\\\hline
e_3^1& 31 & e_6^1 & 46 & e_9^1 & 50\\\hline
\end{array}\]
\item[Step 2] Solve $4500=324x_1-108$
\item [Step 3] $x_1=17$
\end{description}
\item For $B_2$\begin{description}
\item 
[Step1]\[\begin{array}{|c|c|c|c|c|c|}
\hline
e_1^2 & 44 & e_4^2 & 48& e_7^2 & 65 \\\hline
e_2^2 & 27 & e_5^2 & 53  & e_8^2 & 53\\\hline
e_3^2& 54 & e_6^2 & 67 & e_9^2 & 94\\\hline
\end{array}\]
\item[Step 2] Solve $-2710=-243x_2+10071$
\item [Step 3] $x_2=11$
\end{description}
\item For $B_3$ and $B_4$
clearly $x_3=x_4=27.$
\item $M=\begin{bmatrix}27& 4 & 15 & 4 & 26 & 13 \\
8 & 17 & 4 & 17 & 27 & 27\\
27 & 27 & 27 & 27 & 27 & 27 \\
27& 27 & 27& 27 & 27 & 27 \\
27 & 27 & 27 & 27 & 27 & 27 \\
27 & 27 & 27 & 27 & 27 & 27 \\
\end{bmatrix}$
\end{enumerate}
\end{description}

\end{example}
\section{Maple Code}\label{sec:4} 
 with(LinearAlgebra):\\ 
$n := 1:\\ Q := Matrix([[0, 1, 0], [0, 0, 1], [1, 1, 0]]):$\\ 
$Q := Q^n:$\\
$ q[1]:=Q[1,1]:q[2]:=Q[1,2]:q[3]:=Q[1,3]:\\
  q[4]:=Q[2,1]:q[5]:=Q[2,2]:q[6]:=Q[2,3]:\\
  q[7]:=Q[3,1]:q[8]:=Q[3,2]:q[9]:=Q[3,3]:$
\\ 
$B := Matrix([[ b11, b12,b13], [ b21, b22,b23], [b31,b32,b33]]);\#$ the block matrix B\\
$d := Determinant(B);$
 $ b[1] :=B[1,1]:b[2]:=B[1,2]:b[3]:=B[1,3]:\\
  b[4] :=B[2,1]:b[5]:=B[2,2]:b[6]:=B[2,3]:\\
  b[7] :=B[3,1]:b[8]:=B[3,2]:b[9]:=B[3,3]:\\$
$ E :=MatrixMatrixMultiply(Q,B);$
$  e[1] :=E[1,1];e[2]:=q[1]*b[2]+q[3]*b[8];e[3]:=E[1,3];\\
  e[4] :=E[2,1];e[5]:=q[4]*b[2]+q[6]*b[8];e[6]:=E[2,3];\\
  e[7] :=E[3,1];e[8]:=q[7]*b[2]+q[9]*b[8];B[3,2]:e[9]:=E[3,3];$\\
$solve(d=(e[2]+q[2]*x)*Determinant(Matrix([[e[6], e[4]], [e[9],e[7]]]))
\\+ (e[5]+q[5]*x)*Determinant(Matrix([[e[1], e[7]], [e[3],e[9]]]))\\
+(e[8]+q[8]*x)*Determinant(Matrix([[e[3], e[1]], [e[6],e[4]]])),x);$\\

\end{document}